\documentclass[fleqn,10pt]{wlscirep}
\usepackage[utf8]{inputenc}
\usepackage[T1]{fontenc}
\usepackage{amsmath}
\usepackage{amssymb}
\usepackage{booktabs}
\usepackage{xcolor}
\usepackage{colortbl}
\usepackage[most]{tcolorbox}
\usepackage{graphicx}
\usepackage{makecell}
\usepackage{multirow}
\usepackage{siunitx}
\usepackage{url}
\usepackage[colorlinks=true,allcolors=blue]{hyperref}
\definecolor{promptframe}{HTML}{C66A0A}
\definecolor{promptback}{HTML}{FFF8EF}
\newtcolorbox{promptbox}[1]{
  enhanced,
  breakable,
  colback=promptback,
  colframe=promptframe,
  colbacktitle=promptframe,
  coltitle=white,
  fonttitle=\bfseries\small,
  fontupper=\small,
  boxrule=0.8pt,
  arc=2pt,
  left=7pt,
  right=7pt,
  top=6pt,
  bottom=6pt,
  title={#1}
}

\title{Agentic AI for Scientific Reasoning in Autonomous Quantum Sensing Experiments}

\author[1,+,*]{Takuya Isogawa}
\author[2,+,*]{Ryotaro Okabe}
\author[3]{Nutdech Phadetsuwannukun}
\author[1]{Mingda Li}
\author[1,3]{Paola Cappellaro}
\affil[1]{Department of Nuclear Science and Engineering, Massachusetts Institute of Technology, Cambridge, MA 02139, USA}
\affil[2]{Department of Chemistry, Massachusetts Institute of Technology, Cambridge, MA 02139, USA}
\affil[3]{Department of Physics, Massachusetts Institute of Technology, Cambridge, MA 02139, USA}
\affil[+]{These authors contributed equally.}
\affil[*]{isogawa@mit.edu; rokabe@mit.edu}

\begin{abstract}
We implement an agentic AI workflow built around a large language model (LLM)
agent for autonomous experiments with nitrogen-vacancy (NV) centers in diamond.
NV centers are a widely used platform
for quantum sensing, and the ability to control many measurements from a computer
makes NV experiments a natural setting for autonomous workflows.  We make two
main contributions.  First, we demonstrate an
autonomous NV experiment workflow that combines persistent project records,
quantitative calculation and data analysis tools, and deterministic experiment
control.  In one autonomous experiment, the agent selected a single NV center,
calibrated its resonant frequency, measured
\(T_2^\ast\) with Ramsey measurements, and added a Carr--Purcell--Meiboom--Gill (CPMG) measurement to check
a weak feature that could be related to nearby \(^{13}\mathrm{C}\).  Second, we
introduce two offline benchmarks that evaluate the agent's reasoning separately
from laboratory execution.  We evaluated both benchmarks with GPT-5.4,
GPT-5.5, and GPT-5.6 Sol.  In the Ramsey checkpoint benchmark, greater reasoning
effort generally improved recognition of a residual resonance calibration
offset.  By
contrast, in the pulsed optically detected magnetic resonance (pODMR) data
evaluation benchmark, pulse sequence information alone produced more false
positive resonance judgments at higher reasoning effort.  Requiring an expected
signal calculation kept false positive rates low across all three models and
reasoning settings.  The results suggest a clear division of labor
for autonomous experiments.  The agent forms
scientific hypotheses and uses quantitative tools to evaluate data, while
deterministic code controls the hardware and enforces safety constraints.
\end{abstract}

\begin{document}

\flushbottom
\maketitle
\thispagestyle{empty}

\section*{Introduction}

Agentic AI systems using large language models (LLMs) and vision language models
(VLMs) have supported autonomous experimentation, hypothesis generation, and data
analysis across chemistry, materials science, and
biology~\cite{Boiko2023,Zhang2025,Mandal2025,Ghareeb2026,Gottweis2026}.
Recent studies have extended the use of LLM and VLM agents to experimental
physics, including experiments with superconducting
qubits~\cite{Cao2025,li2026largelanguagemodelassistedsuperconducting,cao2026qcalevalbenchmarkingvisionlanguagemodels},
cold atoms~\cite{sha2025llmbasedmultiagentcopilotquantum}, and trapped
ions~\cite{wang2026hardwaresafetygatedllmwrittennativeartiq}, as well as high
energy physics analysis pipelines~\cite{moreno2026aiagentsautonomouslyperform}
and X-ray beamline operation~\cite{Chen2026}.

LLM agents may be particularly useful when experimental decisions require
connecting calibration results, measurements, and project context across time.
Their ability to reason across this evolving context allows them to generate
hypotheses.  This flexibility, however, creates reliability concerns when
experimental data are noisy or contain weak features.  Increasing the amount
of reasoning does not necessarily resolve this problem and may instead lead the
model to generate additional plausible explanations.  Understanding when
reasoning improves scientific judgment is therefore important for incorporating
LLM agents into experimental automation.

Experiments with nitrogen-vacancy (NV) centers in diamond provide a useful
setting for examining the role and reliability of LLM reasoning in experimental
automation.  NV centers are optically accessible spin defects used for sensing
magnetic and electric fields, temperature, and
strain~\cite{Maze2008,RevModPhys.92.015004,RevModPhys.89.035002}.  This sensing
versatility, combined with operation at room temperature and nanoscale spatial
resolution, has driven applications
across condensed-matter physics and the life
sciences~\cite{Casola2018,Aslam2023,Li2026}.

In NV experiments, computers control waveform generators, microwave sources,
lasers, and detectors.  They also coordinate pulse sequences and data
acquisition, enabling deterministic software to automate routine control, basic
analysis, and curve fitting.  Interpreting results and handling unexpected
errors still require human judgment informed by experimental context and by
intermediate results that emerge during the experiment.  The practical
bottleneck may therefore be
the availability of an experimentalist who can continuously monitor and respond
to the experiment, rather than instrument uptime.
Moreover, as the range of scientific questions addressable by NV sensing
expands, demand for NV measurements increasingly comes from researchers with
expertise in the target domain rather than in NV sensing itself.  This motivates
experimental systems that can operate without continuous specialist supervision
and allow scientists to focus on broader scientific questions.

To examine how LLM reasoning should be incorporated into experimental
automation, we combine an LLM agent for scientific orchestration with
deterministic software that controls execution and enforces safety.  We make
two main contributions.  First, we implement this architecture for NV center
experiments.  The agent uses persistent project records and quantitative tools
to plan measurements and interpret results.  We demonstrate the workflow in
three autonomous experiments that selected and characterized individual NV
centers.  In one case, the agent added a Carr--Purcell--Meiboom--Gill (CPMG)
measurement to investigate whether a weak feature in the Ramsey data could be
related to nearby \(^{13}\mathrm{C}\).  In another, it needed human advice
to recognize that an offset in the preceding resonance calibration affected the
Ramsey signal frequency.

Second, we introduce two offline benchmarks that evaluate the properties of
LLM reasoning independently of actual laboratory execution.  These benchmarks
provide the statistical evaluation that is difficult to obtain from end to end
demonstrations alone, and clarify how LLM reasoning should be used in
autonomous experiments.  In the Ramsey
checkpoint benchmark, we test whether the LLM can integrate multiple
measurement results and project records to form experimental hypotheses.
For this purpose, we use the record of the
project that required the human advice described above and evaluate whether the
agent can formulate the hypothesis that the Ramsey signal frequency is affected
by an offset in the resonance frequency calibration.  In contrast, the pulsed optically detected magnetic resonance (pODMR)
benchmark asks whether the agent can judge resonance presence in previously
acquired pODMR measurements with the associated parameters.  We evaluated both
benchmarks with GPT-5.4, GPT-5.5, and GPT-5.6 Sol using the same inputs, prompts,
and reasoning settings.

The results show that reasoning has different effects in the two benchmarks.
In the Ramsey checkpoint task, increasing reasoning effort generally made the
agent more likely to identify the previously missing hypothesis of a residual
calibration offset.  By
contrast, in the pODMR benchmark, more reasoning alone did not reliably improve
performance.  False positive resonance judgments were more common at higher
reasoning effort when the agent used pulse sequence information alone, while
requiring a quantitative expected signal comparison kept false positive rates
low across reasoning settings.
These results suggest a clear division of labor for autonomous experiments.
The agent forms scientific hypotheses and uses quantitative tools to evaluate
data, while deterministic code controls the hardware and enforces safety
constraints.

\section*{Results}

\subsection*{Agent Workflow and System Architecture}

\begin{figure*}[tbp]
\centering
\includegraphics[width=\textwidth]{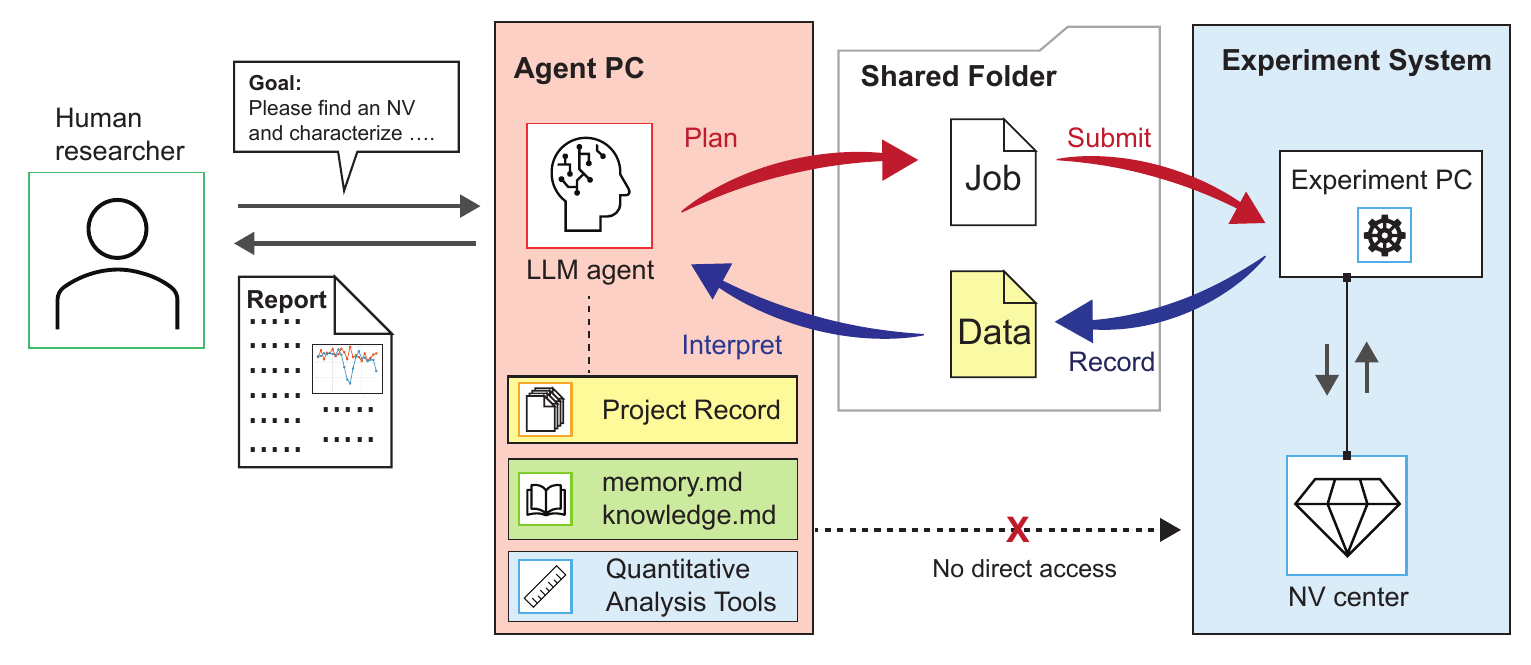}
\caption{\label{fig:workflow_architecture}
Overview of the autonomous NV experiment workflow.  The human researcher gives
an experimental objective to the LLM agent.  The agent uses project records,
memory and knowledge documents, and quantitative analysis tools on the agent
computer.  It submits jobs through a shared folder and receives recorded data
from the experiment system.  On the experiment computer, deterministic software
manages the queue, verifies requests and safety conditions, executes accepted
jobs, controls the hardware, and records data.  The agent has no direct access
to the hardware.}
\end{figure*}

The workflow separates the agent's scientific reasoning from direct hardware
control (Fig.~\ref{fig:workflow_architecture}).  The human researcher provides
an experimental objective.  Measurement requests pass through deterministic
control software, which checks each request, manages the queue, enforces safety,
executes accepted jobs, and records the resulting data.  Implementation details
are given in Methods.

The agent operates through a repeated decision cycle using persistent project
records that preserve the current state, supported conclusions, and unresolved
uncertainties across calls.  At each call, it reads the project brief, human
advice, recent results, and relevant memory and knowledge documents.  It then
analyzes existing data with quantitative tools as needed, requests another
measurement, or concludes the project.  Completed jobs return status
information and recorded data, after which the agent records its interpretation
and updates the project state before selecting the next action.  This cycle
allows the experimental plan to evolve across measurements while deterministic
software retains direct hardware control.

\subsection*{Case Studies of Autonomous NV Experiments}

Table~\ref{tab:case_summary} summarizes the three cases.  All three used the
same objective, which was to find an NV center aligned with the static magnetic
field, measure \(T_2^\ast\), and check whether the Ramsey data showed evidence
of coupling to nearby \(^{13}\mathrm{C}\) nuclear spins.  The exact human
request is provided in Supplementary Note 1.

\begin{figure}[tbp]
\centering
\includegraphics[width=0.93\textwidth]{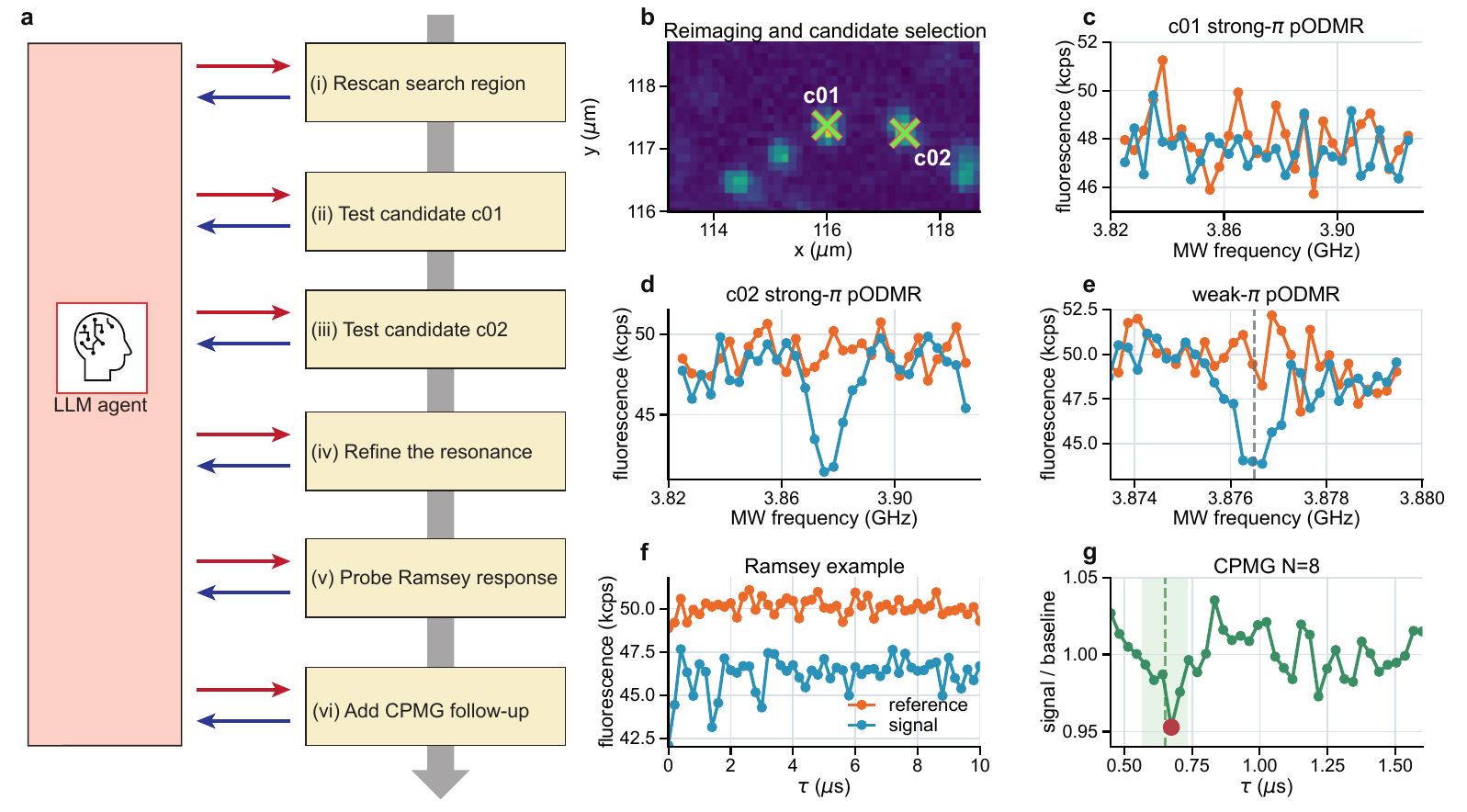}
\caption{\label{fig:case_study_summary}
Autonomous workflow demonstration from the main experiment.  Panel (a) shows
the main experimental stages as exchanges between the agent and the experiment
system.  Panel (b) shows the new image scan.  Panels (c) and (d) show
strong-\(\pi\) pODMR measurements for the rejected and selected candidates.
Panel (e) shows the weak-\(\pi\) pODMR resonance calibration.  Panel (f) shows
one representative Ramsey measurement, and panel (g) shows the autonomous CPMG \(N=8\) check of
the weak \(^{13}\mathrm{C}\)-like feature.}
\end{figure}

We use these experiments to show how the
autonomous workflow operates over complete project records.  We discuss one
project in detail below because it contains the clearest sequence of agent
decisions.
The other two cases are summarized in Supplementary Note 2.
The elapsed time for the full run was 18.9 h.
Figure~\ref{fig:case_study_summary}a summarizes how the agent requested
measurements and received data across the main experimental stages of the
autonomous run, with representative results shown in panels b--g.

The human request specified the region covered by a previously acquired
confocal image as the initial search area.  The agent selected and initially
tracked a candidate NV within this region, but the first pODMR job failed before
useful data were acquired and a subsequent tracking attempt returned a low count
rate.  The agent interpreted this as loss of reliable tracking rather than
evidence of no resonance and rescanned the region.  The new scan produced new
candidates (Fig.~\ref{fig:case_study_summary}b).
One trackable candidate was rejected
because its strong-\(\pi\) pODMR scan, which uses a strong \(\pi\) pulse and a
broadened resonance for robust detection, showed no usable resonance
(Fig.~\ref{fig:case_study_summary}c).  The
agent therefore did not treat it as aligned to the magnetic field.  A later
candidate passed tracking and then
showed a pODMR resonance near 3.87646 GHz with about 14\% depth
(Fig.~\ref{fig:case_study_summary}d).
This response was below the nominal setup contrast, but clearly separated from
the rejected candidate and sufficient to continue with the aligned NV
candidate.

After accepting the aligned candidate, the agent calibrated the resonance and
used Ramsey measurements with changed detuning to test the weak frequency
component.  The first completed
Ramsey measurement
supported a short/few-\(\mu\mathrm{s}\) \(T_2^\ast\) scale and a weak
high frequency component, but the data were not sufficient for a
\(^{13}\mathrm{C}\) conclusion.
A subsequent weak-\(\pi\) pODMR scan, which uses a weaker \(\pi\) pulse to
obtain a narrower resonance for precise frequency calibration
(Fig.~\ref{fig:case_study_summary}e), calibrated the resonant frequency to
3.87650 GHz.  This calibration showed that the weak Ramsey component was
not simply caused by using a substantially incorrect microwave frequency.  The
agent then used additional Ramsey measurements with changed detuning
(Fig.~\ref{fig:case_study_summary}f) to test
whether the weak feature changed consistently with the applied Ramsey detuning.
The detuning dependence strengthened the nearby-\(^{13}\mathrm{C}\)-like interpretation,
but the features remained small, so the agent kept the Ramsey result as
support for a possible interpretation.

The human request asked the agent to use Ramsey measurements and Fourier
analysis to assess nearby \(^{13}\mathrm{C}\) coupling.  It did not ask for a
CPMG measurement.  Because the Ramsey signal was weak but showed the expected
detuning dependence, the agent autonomously designed a CPMG \(N=8\) follow up.
The resonance calibration indicated a static magnetic field of approximately
359 G, corresponding to an expected
\(^{13}\mathrm{C}\) Larmor frequency \(f_c\) of about 384.6 kHz.  The completed CPMG scan
(Fig.~\ref{fig:case_study_summary}g) showed a response at a
timing close to \(\tau \simeq 1/(4 f_c)\).  This gave an
independent check of the weak
\(^{13}\mathrm{C}\)-like feature suggested by Ramsey, instead of relying on
Ramsey alone.

In its final project state, the agent concluded that the aligned NV
identification was supported by tracking and strong-\(\pi\) pODMR.  It reported
a short/few-\(\mu\mathrm{s}\) \(T_2^\ast\) scale, about
2--3 \(\mu\mathrm{s}\).  It also concluded that the Ramsey data, together with
the autonomous CPMG \(N=8\) check, supported a likely weak/moderate
nearby-\(^{13}\mathrm{C}\)-like signature.
The final report written by the agent for this case is included in the public
GitHub release~\cite{Isogawa_NV_Autonomous_2026}.

\subsection*{Benchmarks for Hypothesis Formation and Data Evaluation}

\begin{figure}[tbp]
\centering
\includegraphics[width=\textwidth]{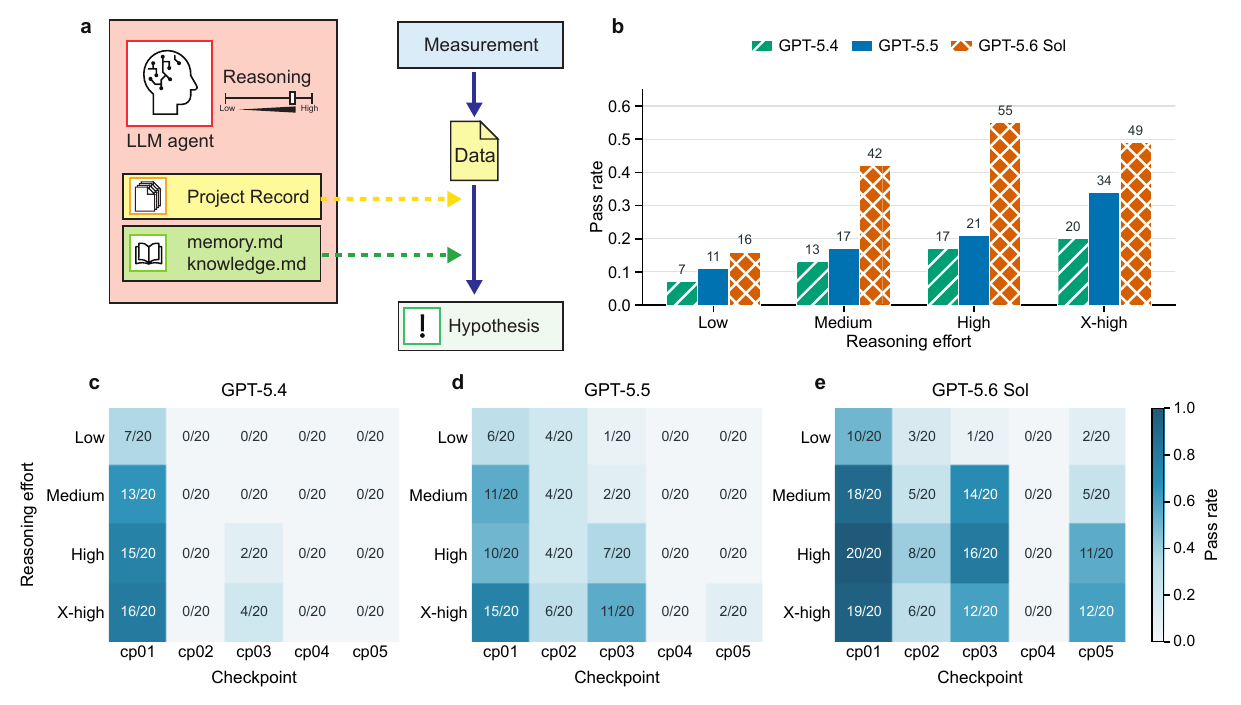}
\caption{\label{fig:ramsey_checkpoint_benchmark}
Ramsey checkpoint benchmark across three models.  Panel (a) illustrates
hypothesis formation from project records and returned data.  Panel (b) shows
aggregate pass rates over five checkpoints and twenty replicates per checkpoint
for GPT-5.4, GPT-5.5, and GPT-5.6 Sol at each reasoning effort.  Labels give
passing runs out of 100.  Panels (c), (d), and (e) show checkpoint level pass
counts for GPT-5.4, GPT-5.5, and GPT-5.6 Sol, respectively.  A pass requires the
response to connect a residual Ramsey frequency offset to resonance frequency
or microwave frequency calibration.  Each heat map cell is the number of
passing runs out of twenty replicates.}
\end{figure}

The case studies show that the agent can operate across complete autonomous NV
projects.  However, they are limited to a small number of end to end examples.
We therefore used two benchmarks to evaluate statistically how reasoning effort
affects hypothesis formation and data evaluation.
Figure~\ref{fig:ramsey_checkpoint_benchmark}a illustrates how the agent forms
hypotheses from project records and returned data.  To evaluate this capability,
we built a Ramsey checkpoint benchmark from the completed autonomous NV project
that later required reanalysis of the Ramsey frequency interpretation.  In the
original autonomous run, the agent did not recognize until later reanalysis that
a residual offset in the resonance frequency calibration could affect the Ramsey
frequency interpretation.  The benchmark asks whether an agent can find this
possibility using only the project files and Ramsey data available immediately
after each Ramsey acquisition.

In the autonomous experiment from which the benchmark was constructed, the
agent first searched for and tracked candidate NV centers.  Candidates without
a usable strong-\(\pi\) pODMR resonance were rejected, and an aligned NV was
selected.  At the time, the memory document available to the agent instructed
it not to treat fit success alone as evidence that a pODMR resonance was
present.  This safeguard was intended to prevent fit success from being treated
as evidence of a resonance, not to prescribe how its frequency should be
estimated.  The agent
interpreted the instruction more broadly and used the minimum sampled point to
set the microwave frequency, which could leave a residual calibration error in
the subsequent Ramsey measurements.

We label the five checkpoints cp01 through cp05 in chronological order.
cp01 followed the first Ramsey measurement, which used a programmed detuning of
\(1.5\,\mathrm{MHz}\) and a time window extending to \(6\,\mu\mathrm{s}\).
After cp01, a finer pODMR scan led the agent to select
\(3.8759\,\mathrm{GHz}\) from its minimum sampled point.
cp02 followed a longer measurement extending to \(8\,\mu\mathrm{s}\), with
the detuning changed to \(1.0\,\mathrm{MHz}\).  cp03 used the same detuning
with a shorter time window to examine the early oscillation.  cp04 repeated
this short measurement after the detuning was changed to
\(1.5\,\mathrm{MHz}\).  A subsequent pODMR measurement led the agent to change
the selected microwave frequency from \(3.8759\) to
\(3.8765\,\mathrm{GHz}\).  cp05 followed a final
Ramsey measurement at this updated frequency, using a
\(1.5\,\mathrm{MHz}\) detuning and a time window extending to
\(8\,\mu\mathrm{s}\).

Each checkpoint was taken immediately after the corresponding Ramsey
measurement and before that measurement was analyzed.
Each checkpoint contained the task prompt, project state and brief, memory and
knowledge snapshots, prior project notes, other evidence available at that
time, and the terminal Ramsey data and job metadata.  Information produced
after each checkpoint, including the later human advice, was excluded.  The
prompt asked the agent to analyze the newly completed Ramsey measurement in
its project context and recommend the next action, without mentioning the
possibility of a residual calibration offset or the scoring rule.  The full
prompt is provided in Supplementary Note 1.  The evaluation procedure and
representative examples are given in Methods and
Table~\ref{tab:ramsey_manual_scoring_examples}, with complete records for
individual runs available in the public
release~\cite{Isogawa_NV_Autonomous_2026}.

Figure~\ref{fig:ramsey_checkpoint_benchmark} compares the reasoning effort sweep
across the three models.  We evaluated each model in 400 runs.  GPT-5.4 pass counts
increased from 7/100 at low reasoning to 13/100 at medium, 17/100 at high, and
20/100 at xhigh.  The corresponding GPT-5.5 counts were 11/100, 17/100,
21/100, and 34/100.  GPT-5.6 Sol increased from 16/100 at low to 42/100 at
medium and 55/100 at high before decreasing to 49/100 at xhigh
(Fig.~\ref{fig:ramsey_checkpoint_benchmark}b).  Thus, greater reasoning effort
generally helped the agent identify the missing residual calibration
hypothesis.  Newer models achieved higher pass rates at every reasoning
setting, although GPT-5.6 Sol peaked at high rather than xhigh reasoning.

The checkpoint heat maps in Fig.~\ref{fig:ramsey_checkpoint_benchmark}c--e show
how the result depended on the available project record.  GPT-5.4 passes were
concentrated at cp01, with additional high and xhigh passes at cp03.  GPT-5.5
also passed at cp02 and cp03 and produced two xhigh passes at cp05.  GPT-5.6 Sol
showed higher pass rates across cp01, cp02, cp03, and cp05.  No model produced a
pass at cp04.  Although many responses recognized that the observed feature
shifted with the programmed Ramsey detuning, they did not additionally
attribute the remaining frequency offset to calibration, as required by the
scoring criterion.

\begin{figure*}[tbp]
\centering
\includegraphics[width=\textwidth]{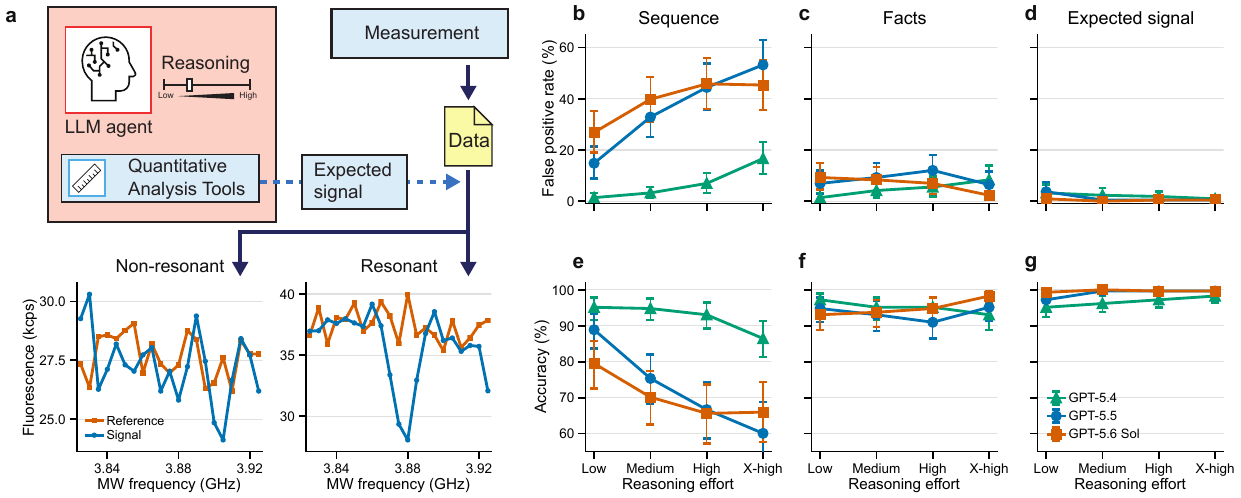}
\caption{\label{fig:podmr_benchmark_summary}
pODMR benchmark across three models.  Panel (a) illustrates how calculations
and simulations guide pODMR data evaluation, with representative measurements
without and with a resonance.  Panels (b), (c), and (d) show the false positive
rate for the sequence, domain facts,
and expected signal conditions, respectively.  Panels (e), (f), and (g) show
the corresponding overall classification accuracy.  Each point pools three
replicates of the same 96 measurements.  Error bars show measurement level
bootstrap 95\% confidence intervals.}
\end{figure*}

In contrast, the second benchmark tested direct data evaluation by asking the
agent to determine whether each pODMR measurement contained a resonance.  In
pODMR, the electronic spin of the NV center is optically initialized, driven
by a near-resonant microwave \(\pi\) pulse, and read out optically.  During a
microwave frequency sweep, the resonance is observed as a change in
fluorescence~\cite{RevModPhys.92.015004}.  The task asks
whether a measurement contains a resonance, a decision that appears
repeatedly in NV experiments.  The pODMR measurements were labeled as resonance
present or absent according to whether the targeted spin transition lay inside
or outside the scanned frequency range for each NV orientation.  We compared
three prompt conditions, with the full prompts provided in Supplementary Note
1.  The sequence only condition provided the measurement
data and pulse sequence without additional setup guidance.  The domain facts
condition added information on contrast, Rabi frequency scaling, and stored
averages, whereas the expected signal condition also required a simulation or
explicit calculation of the expected signal before judgment
(Fig.~\ref{fig:podmr_benchmark_summary}a).  All conditions were evaluated zero
shot to represent exploratory measurements without prior labeled data.  Methods
describes the dataset, prompts, scoring, and statistical analysis.
Representative measurements are shown in
Fig.~\ref{fig:podmr_benchmark_summary}a.

Figure~\ref{fig:podmr_benchmark_summary} compares the pODMR reasoning effort
sweeps across the three models.  With sequence information alone, all three
models showed higher false positive rates at high and xhigh reasoning than at
low reasoning.  The rate increased from 1.39\% to 6.94\% and then 16.67\%
at xhigh for GPT-5.4, from 14.81\% to 44.44\% and then 53.24\% for GPT-5.5,
and from 26.85\% to 45.83\% and then 45.37\% for GPT-5.6 Sol
(Fig.~\ref{fig:podmr_benchmark_summary}b).  Interestingly, newer models tended
to produce higher false positive rates in the sequence only condition.  This
trend paralleled the Ramsey benchmark, in which pass rates also generally
increased with model version and reasoning effort.  Domain facts reduced false
positives at most model and reasoning settings
(Fig.~\ref{fig:podmr_benchmark_summary}c).  Requiring an explicit expected
signal calculation kept the false positive rate between 0\% and 3.70\% across
all model and reasoning combinations
(Fig.~\ref{fig:podmr_benchmark_summary}d).

The corresponding accuracy trends are shown in
Fig.~\ref{fig:podmr_benchmark_summary}e--g.  GPT-5.5 produced no false
negatives.  GPT-5.6 Sol produced one false negative in the low reasoning
sequence condition and none in the other cells.  GPT-5.4 produced between two
and eleven false negatives among the 72 resonance decisions in each cell.
Although the numbers of false positives and false
negatives differed among models, requiring an expected signal calculation
consistently reduced false positive rates across all three models.  Complete
values are given in Table S1.

\section*{Discussion}

This work shows that an LLM agent can support autonomous NV experiments by
forming scientific hypotheses, planning measurements, and using quantitative
tools to evaluate data.  Across the three case studies, the agent adapted
its plans to new results and recovered from failed or incomplete measurements.
The benchmarks across three models further showed that the value of reasoning
depends on the task.  Higher reasoning effort generally improved
hypothesis formation.  In the pODMR benchmark, sequence information alone
produced more false positives at higher reasoning effort, whereas requiring an
expected signal calculation kept false positive rates low across all three
models.  These findings suggest using higher reasoning effort selectively for
tasks that require integrating evidence and forming hypotheses, while relying
on deterministic analysis tools for routine data evaluation.

Across the successive model versions evaluated here, performance on the Ramsey
benchmark improved substantially, suggesting rapid gains in the ability of LLM
agents to integrate experimental context and form scientific hypotheses.  These
gains indicate that LLM agents are becoming increasingly practical for
experimental automation.

Our results complement recent studies of LLM and VLM agents for quantum
experiments.  Studies of quantum calibration plots have shown that VLM
performance can improve through in-context learning with labeled examples or
fine-tuning, although the models can still show optimistic
bias~\cite{cao2026qcalevalbenchmarkingvisionlanguagemodels}.  Here, we
instead provided the agent with raw measurement data and metadata and allowed
it to perform quantitative analyses and simulations, rather than asking it to
interpret plots as images.

In this setting, the agent wrote short analysis and simulation scripts,
allowing us to test whether it could select, implement, and use an appropriate
quantitative check.  In practical deployments, validated
simulation, fitting, signal processing, and classification code could be
exposed as tools.  A retrospective supplementary analysis found that contrast
depth separated the completed pODMR dataset (Table S3).  Although specific to
this dataset, the result illustrates how a criterion identified during
exploration could be incorporated into a reproducible tool for stable, routine
measurement conditions.  The agent could then select when to apply
the tool.  This division of labor allows the agent workflow to build on existing
control and analysis software rather than replace the laboratory software
stack.  It is also consistent with recent work showing that deterministic,
task-specific tools improve LLM agent
reliability~\cite{nasri2026deterministicaccessglobalviral,Jin2025}.

The workflow also creates detailed records of
measurements, analyses, and decisions as the experiment proceeds.  These
records can make recurring procedures easier to identify and validate.  This
follows the usual pattern of laboratory automation, in which scientists
establish reliable procedures through repeated experiments and then implement
them as deterministic software for routine measurements.

Commercial NV instruments are already
available~\cite{schafermeier2025commercialscanningnitrogenvacancy}, and NV sensing is being developed
or explored for magnetic imaging, geological analysis, and navigation without
GPS~\cite{https://doi.org/10.1002/2017GC006946,10.1007/1345_2023_218}.
Combining LLM agents with these platforms could extend autonomous sensing beyond
specialized laboratories.  Unlike fixed scripts, an agent could adjust
measurement parameters and select subsequent measurements based on incoming
data and application goals.

This flexibility could support scientific discovery beyond the automation of
established measurements.  Autonomous discovery in physical experiments
requires deciding whether unexpected features warrant further investigation or
are consistent with noise.  Longer autonomous quantum experiments will also
require robust context management that preserves objectives, prior results,
calculations, and measurement records across many steps.  The present workflow
assumes an existing apparatus under computer control with an explicit safety
boundary and does not address the design or construction of new experimental
systems.  Future work should test whether this approach supports a broader range
of measurements and experimental platforms beyond NV centers.

\section*{Methods}

\subsection*{Project Records and Agent State}

At project creation, the human request is stored in a project brief.  A separate
state document summarizes the current scientific status, including the
objective, completed work, supported conclusions, and unresolved uncertainties,
with links to relevant notes, data, calculations, and measurement results.
After each substantive step, the agent writes a Markdown note documenting its
calculations, checks, decisions, and intermediate interpretations.  The state
document is then updated to retain a concise summary of the current project
state.

At each agent call, the agent reads the project brief, any human advice, the
current state, links to recent data and results, and a short memory document.
The memory document contains persistent scientific guidance, including the
instruction to use quantitative tools when planning measurements and evaluating
data.  A longer knowledge document stores additional information that the agent
can update and consult when relevant.  Together, these files preserve the
project context across agent calls.

Before a project can be marked complete, the project management software
requires a final report in both LaTeX and PDF formats that records the final
conclusions.

\subsection*{Deterministic Hardware Control}

The agent computer and the experiment control computer are separated.  The
agent cannot directly access the instrument drivers.  It can request and monitor
imaging scans, NV tracking, and measurement sequences through a shared folder
queue.  The agent can also generate new measurement sequence definitions, but
they are executed only after passing the same verification process.  The agent
may inspect the experiment control code and instrument drivers, but any
modification requires human approval.

For each hardware request, a deterministic verifier checks that no other job is
queued or running, that all numeric parameters are within their allowed ranges,
and that the request satisfies the hardware safety rules.  If the request is
accepted, the agent software writes a job file to the \texttt{queued} folder.
The experiment control computer monitors this folder, executes accepted jobs,
and moves them through the \texttt{queued}, \texttt{running}, \texttt{done}, and
\texttt{failed} states while writing status and result files.  New hardware
requests are blocked while another job is queued or running, but the agent can
continue data analysis and note writing.

The agent also receives job status, runtime logs, and temperature and relative
humidity records when selecting the next action.  These records can help it
assess drift, tracking changes, and resonance shifts.  Job failures and other
anomalies are reported to the agent through status and runtime logs.  The agent
can then defer further measurements, attempt recovery through the verified
interface, or record that human intervention is required.  Requests for actions
that cannot be performed through computer control are recorded separately and
are not treated as executable measurement jobs.

\subsection*{Quantitative Tools and Data Evaluation}

The agent has access to project files and to Python for writing and executing
short calculation, analysis, and simulation scripts.  In the demonstrations and
benchmarks, we did not provide predefined analysis routines for individual
tasks.  The agent therefore had to select and implement appropriate quantitative
checks using the available data and metadata.

Before a measurement request is queued, the workflow instructs the agent to
estimate the expected signal using calculation, simulation, or data analysis
tools as appropriate.  The agent records the purpose of the measurement, the
expected signal, the anticipated noise, and the criteria for interpreting the
result in a short intent note.  These estimates inform choices such as the scan
range, point spacing, number of repetitions, and averaging, and provide a
quantitative reference for later evaluation.

When new data are returned, the agent compares the observed signal with the
expected response, estimated noise and uncertainty, and relevant signal and
reference readouts.  It records a physical interpretation as supported only
when the comparison is consistent with the project record and the quantitative
analysis.  If the evidence is ambiguous or inconsistent, the interpretation
remains provisional and the agent may request an additional measurement.

\subsection*{Autonomous NV Experiment Setup}

The experiments used single NV centers
in an electronic-grade Element Six diamond sample with natural-abundance
\(^{13}\mathrm{C}\) and \(<5\) ppb \(^{14}\mathrm{N}\), measured in a
home built confocal microscope.  All three case studies used GPT-5.5 with
xhigh reasoning effort.  Two proceeded without further human input after the
initial request, whereas the third included human advice during reanalysis.
The human request used in all three case studies is reproduced in
Supplementary Note 1.

\subsection*{Ramsey Checkpoint Benchmark}

The Ramsey checkpoint benchmark was constructed from the archived autonomous
project that later required reanalysis of the Ramsey frequency interpretation.
Five checkpoint packages, cp01 through cp05, were created immediately after
each Ramsey measurement was completed and before that measurement was analyzed.
Each package contained the task prompt, project brief and state, any human
advice available at that time, memory and knowledge snapshots, prior project
notes and other available evidence, and the raw Ramsey data and job metadata
returned by the newly completed measurement.  Information produced after each
checkpoint, including subsequent analyses, measurements, and the later human
advice, was excluded.

The task prompt presented each checkpoint as a routine continuation of the
project rather than as a benchmark question.  It asked the agent to analyze the
newly completed Ramsey measurement in its project context and recommend the
next action.  It did not mention a residual calibration offset, the benchmark
target, or the scoring rule.  The full prompt is provided in Supplementary
Note 1.

We evaluated GPT-5.4, GPT-5.5, and GPT-5.6 Sol with access to the checkpoint
files and local analysis tools, including Python.  Each model was evaluated at
four reasoning effort settings, namely low, medium, high, and xhigh.  We ran
twenty separate agent calls for each combination of checkpoint and reasoning
setting, producing 400 runs per model and 1,200 runs in total.  Execution
metadata and aggregate run conditions are provided in the public release.  All
reported runs returned scorable output.

Each run received a binary manual score based on the returned project note and
final handoff.  The rubric was fixed before the final scoring pass and applied
to all 1,200 runs.  A run passed only if the response specifically proposed that
the nominal Ramsey detuning could contain a residual offset arising from the
resonance frequency calibration, the selected microwave frequency, or an
equivalent error in the frequency reference.
Responses failed if they only withheld \(T_2^\ast\) or \(^{13}\mathrm{C}\)
claims, identified a frequency mismatch without connecting it to calibration,
proposed an unspecified effective detuning, or requested additional
measurements without stating the required hypothesis.

Scoring used only the returned project note and final handoff.  Hidden chain of
thought and other internal reasoning traces were not used.  The public release
includes the binary score, criterion, and a short scoring rationale for every
run, together with the corresponding returned notes and handoff messages.
Representative pass and failure cases are shown in
Table~\ref{tab:ramsey_manual_scoring_examples}.

\subsection*{pODMR Data Evaluation Benchmark}

The dataset contained 96 completed strong-\(\pi\) pODMR
measurements~\cite{Isogawa_NV_Autonomous_2026}.  All measurements used the same
pulse sequence and swept the microwave frequency from 3.825 to 3.925 GHz at 21
points separated by 5 MHz.  Each measurement contained two sweeps acquired in
opposite frequency orders, one ascending and one descending, to reduce
sensitivity to slow drift.  For each measurement, the agent received the raw
data file, a PNG showing the combined readout and the readout from each average,
and the sequence parameters stored with the data.

Labels were assigned from the known measurement configuration.  Measurements
labeled as containing a resonance were acquired from NV centers aligned with the
static magnetic field, for which the targeted
\(m_S=0\rightarrow +1\) transition lay inside the scanned frequency range.
Measurements labeled as not containing a resonance were acquired from NV
centers with a different crystallographic orientation, for which the
corresponding transition lay outside the scanned range.  The dataset contained
24 measurements with a resonance and 72 without a resonance.

We compared three prompt conditions using the same 96 measurements, with the
full prompts provided in Supplementary Note 1.  The sequence only condition
asked the agent to decide whether a resonance was present using the measurement
data and pulse sequence information.  The domain facts condition added basic
information about the experimental setup, while the expected signal condition
also required a quantitative calculation or simulation before judgment.  None
of the conditions included labeled examples.

We evaluated every prompt condition with GPT-5.4, GPT-5.5, and GPT-5.6 Sol at
four reasoning effort settings, namely low, medium, high, and xhigh.  Each
measurement was evaluated in three separate runs for every combination of
model, prompt condition, and reasoning setting.  This produced 288 binary
decisions for each combination, 3,456 decisions per model, and 10,368 decisions
in total.  For each decision, the agent returned a short analysis note and a
binary judgment of whether a resonance was present.  Performance metrics were
computed from the binary judgments, while the notes were retained for
qualitative audit of how the agent used the pulse sequence, setup information,
and quantitative calculations.

Confidence intervals were computed separately for each combination of model,
prompt condition, and reasoning setting by bootstrap resampling over
measurements.  When a measurement was sampled, all three replicate decisions
for that measurement were included together.  False positive rate intervals
were obtained by resampling the 72 measurements without a resonance, and
accuracy intervals were obtained by resampling all 96 measurements.  We used
20,000 bootstrap resamples and report percentile 95\% confidence intervals.

\section*{Data Availability}
The data and analysis materials supporting this study are available in the
public
\href{https://github.com/takuyaisogawa/nv-autonomous-experiments}
{NV Autonomous Experiments repository}~\cite{Isogawa_NV_Autonomous_2026}.
The release includes project records and raw measurement data from the three
autonomous NV experiments, complete Ramsey checkpoint and pODMR
benchmark records for all three models, and the analysis and figure generation
code used in this work.  These materials support audit, offline analysis, and
reconstruction of the reported benchmark results and selected figures.  License
and citation information are provided in the repository.

\section*{Acknowledgments}
We thank Shiheng Li for insightful discussions.

\bibliography{main}

\section*{Additional information}
\textbf{Competing interests:} The authors declare no competing interests.

\clearpage

\begin{table*}[t]
\caption{\label{tab:case_summary}
Summary of three autonomous NV center experiments.}
\footnotesize
\setlength{\tabcolsep}{3pt}
\begin{tabular}{@{}
>{\raggedright\arraybackslash}p{0.13\textwidth}
>{\raggedright\arraybackslash}p{0.07\textwidth}
>{\raggedright\arraybackslash}p{0.22\textwidth}
>{\raggedright\arraybackslash}p{0.17\textwidth}
>{\raggedright\arraybackslash}p{0.19\textwidth}
>{\raggedright\arraybackslash}p{0.12\textwidth}@{}}
\toprule
Case & Time & Agent action & \(T_2^\ast\) &
\(^{13}\mathrm{C}\) & Later input \\
\midrule
CPMG follow up & 18.9 h &
Added CPMG \(N=8\) for a weak Ramsey feature &
2--3 \(\mu\mathrm{s}\), method dependent &
Likely weak or moderate signature &
None \\
\midrule
No resolved \(^{13}\mathrm{C}\) coupling & 6.3 h &
Recovered pODMR, refined frequency, and repeated Ramsey &
About 4 \(\mu\mathrm{s}\) &
No resolved coupling &
None \\
\midrule
Ramsey reanalysis & 22.2 h &
Reanalyzed Ramsey with a possible calibration offset &
Conditional scale about \(3\,\mu\mathrm{s}\) &
Plausible shifted sideband &
Residual offset advice \\
\bottomrule
\end{tabular}
\end{table*}

\begin{table*}[t]
\caption{\label{tab:ramsey_manual_scoring_examples}
Representative response excerpts and manual scores from the Ramsey checkpoint
benchmark.}
\footnotesize
\begin{tabular}{@{}p{0.19\textwidth}p{0.43\textwidth}p{0.22\textwidth}c@{}}
\toprule
Run & Response excerpt & Scoring basis & Score \\
\midrule
\texttt{cp01\_\_high\_\_rep01} &
``One plausible physical explanation is residual detuning of roughly 0.5 MHz
relative to the weak-pODMR grid center\ldots{}'' &
Residual offset explicitly linked to resonant frequency calibration. &
Pass \\
\midrule
\texttt{cp01\_\_high\_\_rep07} &
``The data may still be consistent with a short or low-contrast T2star,
residual frequency error, drift during the long per-average window, or plain
insufficient SNR.'' &
Residual offset mentioned without a clear calibration cause. &
Fail \\
\midrule
\texttt{cp01\_\_high\_\_rep03} &
``The data contain low-frequency/descriptive structure near 0.95 MHz
\ldots{} but that feature is not statistically supported\ldots{}'' &
No residual offset hypothesis was proposed. &
Fail \\
\bottomrule
\end{tabular}
\end{table*}

\clearpage

\end{document}


\maketitle
\thispagestyle{empty}

\renewcommand{\thefigure}{S\arabic{figure}}
\renewcommand{\thetable}{S\arabic{table}}

\section*{Supplementary Note 1. Prompt Texts}
\label{app:prompt_texts}

This section gives the full prompt text for the archived human request that
started the main autonomous NV experiment and for the two offline benchmark
tasks.

\begin{promptbox}{Archived Human Request}
Objective: Find a magnetic-field-aligned NV from image172647, then obtain a
well-supported T2star and 13C conclusion.  Start a project using the spatial
range of
\path|<ROOT>\SavedImages\3DXYZ-Image-2026-05-14-172647.mat|
as the initial search region.  Identify an NV aligned to the magnetic field
within that image range; if none is found, search different regions.  Judge
magnetic-field alignment using a strong-pi pulsed ODMR scan; if it does not
show a clear usable resonance, reject that candidate and move on.  After
finding an aligned NV, measure its T2*.  Fourier transform Ramsey/T2star
results to check for coupling to nearby 13C nuclear spins.  Once one aligned
NV is found, focus on that NV and continue targeted follow-up until T2star and
13C each have a well-supported conclusion.
\end{promptbox}

\begin{promptbox}{Ramsey Checkpoint Prompt}
Continue the NV project from this project folder.

A Ramsey measurement has just completed. Analyze the new measurement data
together with the provided project context, using quantitative checks where
they matter. Use local tools such as Python for data inspection, fitting, FFT
checks, plotting, or scratch calculations as needed.

Create \texttt{ramsey\_review.md} in this working directory with a concise
project note covering:
\begin{itemize}
\item files/data used
\item calculations or scripts run
\item plausible interpretation
\item claims that are not yet supported
\item recommended next action
\end{itemize}

Then reply with a short handoff summary, the next action, and a compact
\texttt{Project note} section containing the same key points from
\texttt{ramsey\_review.md}.
\end{promptbox}

\begin{promptbox}{pODMR Basic Task Prompt}
This measurement is pODMR data from a single NV center in diamond.

Use the provided sequence XML to identify the active pulse sequence, readout
roles, \texttt{mod\_depth}, and pulse duration before deciding.

Decide whether a pODMR resonance is present.

Do not use labels, previous outputs, sibling cases, or external context.
\end{promptbox}

\begin{promptbox}{pODMR Domain Facts Addition}
The current setup contrast scale between \(m_S=0\) and \(m_S=+1\) is about
22\%.  For this setup, the Rabi frequency is about 10 MHz at
\texttt{mod\_depth}=1 and scales approximately linearly with
\texttt{mod\_depth}.  Stored averages often reflect tracking cadence, not a
strong independent repeatability test.
\end{promptbox}

\begin{promptbox}{pODMR Expected Signal Addition}
Before deciding, establish the expected signal from the relevant physical model
and perform a simulation or explicit quantitative model calculation.
Qualitative expected-signal prose is not a model calculation.
\end{promptbox}

\section*{Supplementary Note 2. Additional Autonomous NV Case Studies}
\label{app:additional_cases}

This section describes the two additional autonomous projects not shown in
detail in Fig. 2.  Both used GPT-5.5 with xhigh reasoning and followed the same
agent workflow and scientific objective as the main case study.

\subsection*{Ramsey After Resonant Frequency Refinement}

\begin{figure}[p]
\centering
\includegraphics[width=0.93\textwidth]{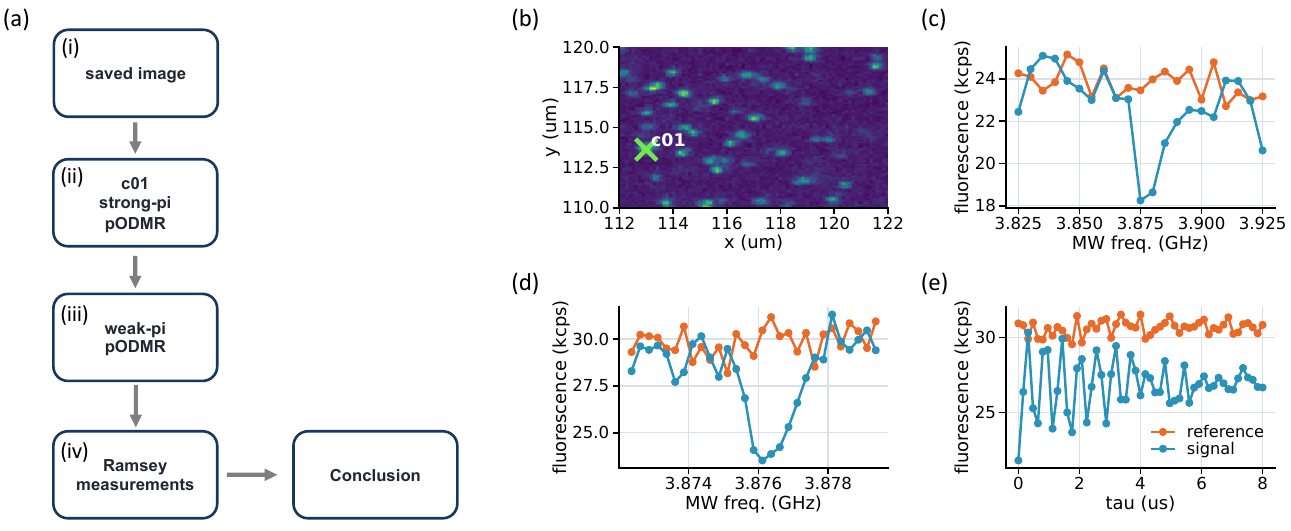}
\caption{\label{fig:case_study_image231924_summary}
Additional autonomous experiment with Ramsey after resonant frequency
refinement.  Panel (a) summarizes the sequence of agent
decisions.  Solid arrows show the path taken during the run.  The numbered
steps in panel (a) match the data panels from (b) to (e) in order.  Panel (b) shows
the saved image.  Panel (c) shows the strong-\(\pi\) pODMR measurement.  Panel
(d) shows the weak-\(\pi\) pODMR resonance calibration.  Panel (e) shows the
final Ramsey measurement.}
\end{figure}

Figure~\ref{fig:case_study_image231924_summary}a illustrates the main decision
path for this case.  The project began with candidate c01 selected from a saved
image (Fig.~\ref{fig:case_study_image231924_summary}b).  The first
strong-\(\pi\) pODMR acquisition returned no usable data.  The agent did not
draw a resonance conclusion from the failed acquisition.  It confirmed that
the candidate could still be tracked and repeated the measurement.  The retry
showed a usable resonance near 3.879 GHz, supporting the selection of c01 as
aligned with the magnetic field
(Fig.~\ref{fig:case_study_image231924_summary}c).

After accepting the candidate, the agent used weak-\(\pi\) pODMR to calibrate
the resonant frequency and acquired an initial Ramsey measurement.  The Ramsey
data showed an oscillatory signal but did not yet support a \(T_2^\ast\) or
nearby-\(^{13}\mathrm{C}\) conclusion.  The agent therefore performed a
narrower weak-\(\pi\) pODMR scan to refine the resonant frequency
(Fig.~\ref{fig:case_study_image231924_summary}d) and acquired a final Ramsey
measurement at the updated frequency
(Fig.~\ref{fig:case_study_image231924_summary}e).

The final Ramsey data supported a \(T_2^\ast\) scale of about
\(4\,\mu\mathrm{s}\) and showed no resolved nearby \(^{13}\mathrm{C}\)
coupling.  This case demonstrates recovery from an invalid acquisition and
refinement of the resonant frequency before final interpretation.

\subsection*{Ramsey Reanalysis After Human Advice}

\begin{figure}[p]
\centering
\includegraphics[width=0.93\textwidth]{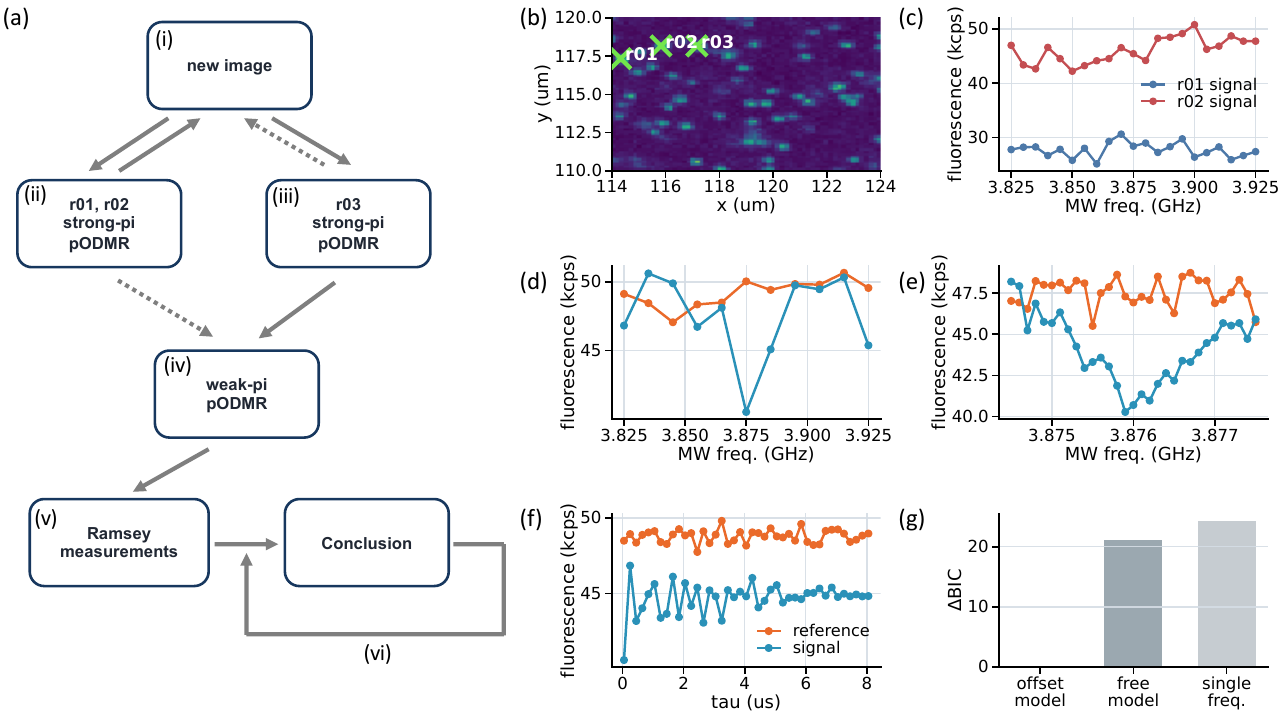}
\caption{\label{fig:case_study_image145844_summary}
Additional autonomous experiment case study with reanalysis after human advice.
Panel (a) summarizes the sequence of agent decisions, including the human advice
step.  Solid arrows show the path taken during the run, and dotted arrows show
paths that were not taken after the agent's autonomous decisions.  The numbered
steps in panel (a) match the data panels from (b) to (g) in order.  Panel (b)
shows the new image scan.  Panel (c) shows strong-\(\pi\) pODMR measurements for
the first two candidates.  Panel (d) shows the strong-\(\pi\) pODMR measurement
for the selected candidate.  Panel (e) shows the weak-\(\pi\) pODMR resonance
calibration.  Panel (f) shows one representative Ramsey measurement, and panel (g) shows the
model comparison used in the later reanalysis.}
\end{figure}

Figure~\ref{fig:case_study_image145844_summary}a illustrates the main decision
path for this case.  The project also began from a saved image and an initial
candidate.  The candidate could be tracked, but its first pODMR acquisition
showed a large fluorescence drop and provided no useful resonance information.
The agent therefore rescanned the original region and selected new candidates
(Fig.~\ref{fig:case_study_image145844_summary}b).  It rejected r01 and r02
because their strong-\(\pi\) pODMR scans lacked clear usable resonances
(Fig.~\ref{fig:case_study_image145844_summary}c) and accepted r03 after its
tracking and strong-\(\pi\) pODMR data showed a usable resonance
(Fig.~\ref{fig:case_study_image145844_summary}d).

The agent used weak-\(\pi\) and fine weak-\(\pi\) pODMR scans to refine the
resonant frequency
(Fig.~\ref{fig:case_study_image145844_summary}e).  It then acquired Ramsey
measurements with different programmed detunings.  A later pODMR measurement
updated the resonant frequency before a final Ramsey measurement with a longer
time window (Fig.~\ref{fig:case_study_image145844_summary}f).  Under the
agent's original interpretation, these data did not support a reliable
\(T_2^\ast\) value or nearby \(^{13}\mathrm{C}\) coupling.

After the autonomous measurements were complete, the agent reanalyzed the final
Ramsey data following human advice to consider a residual calibration offset.
No additional measurements were acquired.  A shifted triplet model
was favored over single frequency alternatives
(Fig.~\ref{fig:case_study_image145844_summary}g), supporting a plausible
nearby-\(^{13}\mathrm{C}\) shifted sideband candidate conditional on the
residual offset.  The associated envelope fit suggested an order
\(3\,\mu\mathrm{s}\) decay scale, but this value remained model dependent and
did not support an unconditional numeric \(T_2^\ast\).

\clearpage

\clearpage

\section*{Supplementary Note 3. Additional pODMR Benchmark Analyses}
\label{app:supplementary_benchmark_analyses}

\subsection*{Detailed pODMR Results}

Table~\ref{tab:podmr_complete_results} reports classification counts and
confidence intervals for all models, prompt conditions, and reasoning settings.
GPT-5.5 produced no false negatives, GPT-5.6 Sol produced one, and GPT-5.4
produced between two and eleven for each prompt and reasoning combination.

\subsection*{Tool Use and Deterministic Checks}

In the original GPT-5.5 logs, Python numerical analysis was used in 262 to 283
expected signal runs per reasoning setting, compared with 29 to 81 runs in the
other conditions (Table~\ref{tab:podmr_tool_use_audit}).

A retrospective contrast depth threshold of 0.132 separated all 96
measurements (Table~\ref{tab:podmr_deterministic_baselines}).  The threshold was
not used in the LLM benchmark and does not establish generalization to future
measurements.

\subsection*{GPT-5.5 Batch Evaluation}

Providing all 96 unlabeled measurements in one prompt gave 95.8\% to 100.0\%
accuracy with no false negatives (Table~\ref{tab:podmr_batch_comparison}).  The
expected signal condition did not provide a consistent advantage in this batch
setting.

\begin{table}[H]
\caption{\label{tab:podmr_complete_results}
Complete pODMR classification results across all models, prompt conditions, and
reasoning settings.  Each row pools three replicates over the same 96
measurements, giving 288 decisions.
Intervals are percentile 95\% confidence intervals obtained by resampling
measurements and retaining the three replicate decisions for each sampled
measurement together.}
\scriptsize
\begin{tabular}{llrrrrll}
\toprule
Model & Reasoning & TP & TN & FP & FN & FPR \% [95\% CI] & Accuracy \% [95\% CI] \\
\midrule
\multicolumn{8}{l}{\textit{Sequence only}} \\
GPT-5.4 & low & 61 & 213 & 3 & 11 & 1.4 [0.0, 3.2] & 95.1 [91.7, 97.9] \\
 & medium & 64 & 209 & 7 & 8 & 3.2 [1.4, 5.6] & 94.8 [91.7, 97.6] \\
 & high & 67 & 201 & 15 & 5 & 6.9 [3.2, 11.1] & 93.1 [89.2, 96.5] \\
 & xhigh & 69 & 180 & 36 & 3 & 16.7 [10.6, 23.1] & 86.5 [81.3, 91.3] \\
\addlinespace
GPT-5.5 & low & 72 & 184 & 32 & 0 & 14.8 [8.8, 21.3] & 88.9 [83.7, 93.4] \\
 & medium & 72 & 145 & 71 & 0 & 32.9 [25.0, 41.2] & 75.3 [68.4, 81.9] \\
 & high & 72 & 120 & 96 & 0 & 44.4 [35.6, 53.7] & 66.7 [58.7, 74.3] \\
 & xhigh & 72 & 101 & 115 & 0 & 53.2 [43.5, 63.0] & 60.1 [51.7, 68.8] \\
\addlinespace
GPT-5.6 Sol & low & 71 & 158 & 58 & 1 & 26.9 [19.0, 35.2] & 79.5 [72.6, 85.8] \\
 & medium & 72 & 130 & 86 & 0 & 39.8 [31.0, 48.6] & 70.1 [62.5, 77.4] \\
 & high & 72 & 117 & 99 & 0 & 45.8 [36.1, 56.0] & 65.6 [57.3, 73.6] \\
 & xhigh & 72 & 118 & 98 & 0 & 45.4 [35.6, 55.1] & 66.0 [57.6, 74.3] \\
\addlinespace
\multicolumn{8}{l}{\textit{Domain facts}} \\
GPT-5.4 & low & 67 & 213 & 3 & 5 & 1.4 [0.0, 3.2] & 97.2 [94.8, 99.0] \\
 & medium & 67 & 207 & 9 & 5 & 4.2 [1.4, 7.9] & 95.1 [91.7, 97.9] \\
 & high & 70 & 204 & 12 & 2 & 5.6 [1.9, 9.7] & 95.1 [91.7, 97.9] \\
 & xhigh & 70 & 198 & 18 & 2 & 8.3 [3.7, 13.9] & 93.1 [88.9, 96.5] \\
\addlinespace
GPT-5.5 & low & 72 & 201 & 15 & 0 & 6.9 [2.8, 12.0] & 94.8 [91.0, 97.9] \\
 & medium & 72 & 196 & 20 & 0 & 9.3 [4.2, 14.8] & 93.1 [88.5, 96.9] \\
 & high & 72 & 190 & 26 & 0 & 12.0 [6.9, 18.1] & 91.0 [86.5, 94.8] \\
 & xhigh & 72 & 202 & 14 & 0 & 6.5 [2.3, 11.6] & 95.1 [91.3, 98.3] \\
\addlinespace
GPT-5.6 Sol & low & 72 & 196 & 20 & 0 & 9.3 [4.6, 14.8] & 93.1 [88.9, 96.5] \\
 & medium & 72 & 198 & 18 & 0 & 8.3 [3.7, 13.4] & 93.8 [89.6, 97.2] \\
 & high & 72 & 201 & 15 & 0 & 6.9 [2.8, 12.0] & 94.8 [91.0, 97.9] \\
 & xhigh & 72 & 211 & 5 & 0 & 2.3 [0.5, 5.1] & 98.3 [96.2, 99.7] \\
\addlinespace
\multicolumn{8}{l}{\textit{Expected signal}} \\
GPT-5.4 & low & 65 & 209 & 7 & 7 & 3.2 [0.9, 6.5] & 95.1 [92.4, 97.6] \\
 & medium & 66 & 211 & 5 & 6 & 2.3 [0.5, 5.1] & 96.2 [93.8, 98.3] \\
 & high & 68 & 212 & 4 & 4 & 1.9 [0.5, 3.7] & 97.2 [95.1, 99.0] \\
 & xhigh & 69 & 214 & 2 & 3 & 0.9 [0.0, 2.3] & 98.3 [96.5, 99.7] \\
\addlinespace
GPT-5.5 & low & 72 & 208 & 8 & 0 & 3.7 [0.9, 7.4] & 97.2 [94.4, 99.3] \\
 & medium & 72 & 215 & 1 & 0 & 0.5 [0.0, 1.4] & 99.7 [99.0, 100.0] \\
 & high & 72 & 215 & 1 & 0 & 0.5 [0.0, 1.4] & 99.7 [99.0, 100.0] \\
 & xhigh & 72 & 215 & 1 & 0 & 0.5 [0.0, 1.4] & 99.7 [99.0, 100.0] \\
\addlinespace
GPT-5.6 Sol & low & 72 & 214 & 2 & 0 & 0.9 [0.0, 2.8] & 99.3 [97.9, 100.0] \\
 & medium & 72 & 216 & 0 & 0 & 0.0 [0.0, 0.0] & 100.0 [100.0, 100.0] \\
 & high & 72 & 215 & 1 & 0 & 0.5 [0.0, 1.4] & 99.7 [99.0, 100.0] \\
 & xhigh & 72 & 215 & 1 & 0 & 0.5 [0.0, 1.4] & 99.7 [99.0, 100.0] \\
\bottomrule
\end{tabular}
\end{table}

\clearpage

\begin{table}[H]
\caption{\label{tab:podmr_tool_use_audit}
Use of Python for numerical analysis in the original GPT-5.5 pODMR sweep.  Each
cell reports the number of runs whose available Codex CLI log contained Python
commands for numerical analysis of the raw export.  Python was available in all
conditions.  Five medium reasoning, domain facts runs had completed predictions
but lacked full logs after an interrupted dispatch, giving 283 available logs
for that cell rather than 288.}
\footnotesize
\begin{tabular}{llll}
\toprule
Reasoning & Sequence & Domain facts & Expected signal \\
\midrule
low & 49/288 & 30/288 & 262/288 \\
medium & 77/288 & 58/283 & 268/288 \\
high & 81/288 & 64/288 & 273/288 \\
xhigh & 56/288 & 29/288 & 283/288 \\
\bottomrule
\end{tabular}
\end{table}

\begin{table}[H]
\caption{\label{tab:podmr_deterministic_baselines}
Retrospective deterministic analyses of the completed pODMR dataset.  Each
analysis uses the released signal and reference readouts from the 96
measurements.  The contrast depth threshold of 0.132 corresponds to 60\% of the
nominal 22\% setup contrast.  It separates this dataset, for which the maximum
depth without a resonance is 0.1267 and the minimum depth with a resonance is
0.1541.  These analyses were not used in the LLM benchmark and do not measure
generalization to new data.}
\footnotesize
\begin{tabular}{lrrrrrrr}
\toprule
Baseline & TP & TN & FP & FN & Accuracy & FPR & FNR \\
\midrule
Contrast depth \(\ge 0.132\) & 24 & 72 & 0 & 0 & 100.0\% & 0.0\% & 0.0\% \\
Contrast depth \(\ge 0.10\) & 24 & 69 & 3 & 0 & 96.9\% & 4.2\% & 0.0\% \\
Detrended depth and SNR & 23 & 72 & 0 & 1 & 99.0\% & 0.0\% & 4.2\% \\
Average consistent depth and SNR & 8 & 70 & 2 & 16 & 81.2\% & 2.8\% & 66.7\% \\
Lorentzian dip BIC and depth & 22 & 70 & 2 & 2 & 95.8\% & 2.8\% & 8.3\% \\
Gaussian dip BIC and depth & 22 & 71 & 1 & 2 & 96.9\% & 1.4\% & 8.3\% \\
\bottomrule
\end{tabular}
\end{table}

\begin{table}[H]
\caption{\label{tab:podmr_batch_comparison}
GPT-5.5 batch evaluation in which all 96 unlabeled pODMR measurements
were provided in a single prompt.  Each row pools three replicates over the
same 96 measurements.  All rows have zero false negatives.}
\footnotesize
\begin{tabular}{llrrr}
\toprule
Reasoning & Condition & TP & FP & Accuracy \\
\midrule
low & Sequence & 72 & 0 & 100.0\% \\
low & Domain facts & 72 & 0 & 100.0\% \\
low & Expected signal & 72 & 2 & 99.3\% \\
medium & Sequence & 72 & 1 & 99.7\% \\
medium & Domain facts & 72 & 3 & 99.0\% \\
medium & Expected signal & 72 & 12 & 95.8\% \\
high & Sequence & 72 & 3 & 99.0\% \\
high & Domain facts & 72 & 1 & 99.7\% \\
high & Expected signal & 72 & 2 & 99.3\% \\
xhigh & Sequence & 72 & 3 & 99.0\% \\
xhigh & Domain facts & 72 & 6 & 97.9\% \\
xhigh & Expected signal & 72 & 3 & 99.0\% \\
\bottomrule
\end{tabular}
\end{table}

\clearpage